\documentclass[12pt,english]{article}
\usepackage{enumitem}
\usepackage{lscape}
\usepackage[utf8]{inputenc}
\usepackage{graphicx}
\usepackage[titletoc,title]{appendix}
\usepackage{amsmath,amssymb,bm}
\usepackage[hidelinks]{hyperref}
\usepackage[export]{adjustbox}[2011/08/13]
\usepackage{chngcntr}
\usepackage{gensymb}

\usepackage[backend=biber,style=apa,natbib=true,sortcites=false]{biblatex}
\allowdisplaybreaks
\usepackage{makecell}
\usepackage{xcolor}
\usepackage{upgreek}
\usepackage{siunitx}
\usepackage{lineno}
\usepackage{yfonts}

\usepackage{booktabs,threeparttable}

\usepackage{multirow}
\usepackage{booktabs}
\usepackage{url}
\usepackage{cancel}
\usepackage{soul} 
 
\usepackage{setspace}
\usepackage[hang]{footmisc}
\usepackage{algorithm}
\usepackage{algpseudocode}
\algrenewcommand\textproc{}

\renewcommand{\algorithmicensure}{\textbf{Output:}}
\usepackage{calc} 
\algrenewcommand\algorithmicensure{%
  \makebox[\widthof{\textbf{Require:}}][l]{\textbf{Ensure:}}}
\let\oldReturn\Return
\renewcommand{\Return}{\State\oldReturn}
\algdef{SE}{Begin}{End}{\textbf{begin}}{\textbf{end}}
\makeatletter
\newcommand*{\algrule}[1][\algorithmicindent]{\makebox[#1][l]{\hspace*{.5em}\thealgruleextra\vrule height \thealgruleheight depth \thealgruledepth}}%
\newcommand*{\thealgruleextra}{}
\newcommand*{\thealgruleheight}{.75\baselineskip}
\newcommand*{\thealgruledepth}{.25\baselineskip}

\newcount\ALG@printindent@tempcnta
\def\ALG@printindent{%
    \ifnum \theALG@nested>0
        \ifx\ALG@text\ALG@x@notext
        \else
            \unskip
            \addvspace{-1pt}
            \ALG@printindent@tempcnta=1
            \loop
                \algrule[\csname ALG@ind@\the\ALG@printindent@tempcnta\endcsname]%
                \advance \ALG@printindent@tempcnta 1
            \ifnum \ALG@printindent@tempcnta<\numexpr\theALG@nested+1\relax
            \repeat
        \fi
    \fi
    }%
\usepackage{etoolbox}
\patchcmd{\ALG@doentity}{\noindent\hskip\ALG@tlm}{\ALG@printindent}{}{\errmessage{failed to patch}}

\newbox\statebox
\newcommand{\myState}[1]{%
    \setbox\statebox=\vbox{#1}%
    \edef\thealgruleheight{\dimexpr \the\ht\statebox+1pt\relax}%
    \edef\thealgruledepth{\dimexpr \the\dp\statebox+1pt\relax}%
    \ifdim\thealgruleheight<.75\baselineskip
        \def\thealgruleheight{\dimexpr .75\baselineskip+1pt\relax}%
    \fi
    \ifdim\thealgruledepth<.25\baselineskip
        \def\thealgruledepth{\dimexpr .25\baselineskip+1pt\relax}%
    \fi
    \State #1%
    \def\thealgruleheight{\dimexpr .75\baselineskip+1pt\relax}%
    \def\thealgruledepth{\dimexpr .25\baselineskip+1pt\relax}%
}

\usepackage[a4paper, total={7in, 9in}]{geometry}
\usepackage[labelfont=bf]{caption}
\graphicspath{ {./images/} }

\usepackage{csquotes}
\usepackage[english]{babel}
\usepackage{comment}
\makeatletter
\def\@biblabel#1{}
\makeatother

\usepackage{wrapfig}
\usepackage{rotating}
\usepackage{epstopdf}

\usepackage{array}
\newcolumntype{L}[1]{>{\raggedright\let\newline\\\arraybackslash\hspace{0pt}}m{#1}}
\newcolumntype{C}[1]{>{\centering\let\newline\\\arraybackslash\hspace{0pt}}m{#1}}
\newcolumntype{R}[1]{>{\raggedleft\let\newline\\\arraybackslash\hspace{0pt}}m{#1}}

\newcommand*\patchAmsMathEnvironmentForLineno[1]{
  \expandafter\let\csname old#1\expandafter\endcsname\csname #1\endcsname
  \expandafter\let\csname oldend#1\expandafter\endcsname\csname end#1\endcsname
  \renewenvironment{#1}
  {\linenomath\csname old#1\endcsname}
  {\csname oldend#1\endcsname\endlinenomath}}
  \newcommand*\patchBothAmsMathEnvironmentsForLineno[1]{
  \patchAmsMathEnvironmentForLineno{#1}
  \patchAmsMathEnvironmentForLineno{#1*}}
  \AtBeginDocument{
  \patchBothAmsMathEnvironmentsForLineno{equation}
  \patchBothAmsMathEnvironmentsForLineno{align}
  \patchBothAmsMathEnvironmentsForLineno{flalign}
  \patchBothAmsMathEnvironmentsForLineno{alignat}
  \patchBothAmsMathEnvironmentsForLineno{gather}
  \patchBothAmsMathEnvironmentsForLineno{multline}
}

\usepackage{xcite}
\usepackage{xr}
\makeatletter
\newcommand*{\addFileDependency}[1]{
  \typeout{(#1)}
  \@addtofilelist{#1}
  \IfFileExists{#1}{}{\typeout{No file #1.}}
}
\makeatother

\usepackage{mathtools}

\begin{document}

\setlength\abovedisplayskip{4pt}
\setlength\belowdisplayskip{4pt}

\title{Graph-Learned Spring SST Evolution and Land-Ocean Synergy Improve Yangtze Summer Rainfall Prediction}

\author{Yifu Gao\footnotemark[1]\;\,\footnotemark[2]\;, 
        Haishan Chen\footnotemark[1]\;\,\footnotemark[2]\;\,\footnotemark[5], 
        Runze Li\footnotemark[3]\;,
        Efi Foufoula-Georgiou\footnotemark[3]\;\footnotemark[4]\;,
        and 
        Jasper A. Vrugt\footnotemark[3]\;\footnotemark[4]\;}

\footnotetext[1]{State Key Laboratory of Climate System Prediction and Risk Management/Key Laboratory of Meteorological Disaster, Ministry of Education/Collaborative Innovation Center on Forecast and Evaluation of Meteorological Disasters, Nanjing University of Information Science and Technology, Nanjing, Jiangsu, China.}
\footnotetext[2]{School of Atmospheric Sciences, Nanjing University of Information Science and Technology, Nanjing, Jiangsu, China.}
\footnotetext[3]{Department of Civil and Environmental Engineering, University of California, Irvine, California, USA.}
\footnotetext[4]{Department of Earth System Science, University of California, Irvine, Irvine, California, USA.}
\footnotetext[5]{Corresponding Author: \textcolor[rgb]{0.00,0.00,1.00}{haishan@nuist.edu.cn}}

\maketitle

\newpage

\begin{center}
\textbf{\large{Abstract}}
\end{center}

Seasonal prediction of Yangtze summer rainfall is challenged by nonstationary oceanic precursors. Here we demonstrate that accounting for the evolution of spring SST and state-dependent land surface information improves predictability. Graph total variation (GTV) is applied to March-May SST fields over the western Pacific, Ni\~{n}o3.4, and tropical Indian Ocean to identify spatially coherent, month-specific SST coefficients for each grid cell and month to form a scalar GTV index. The learned coefficient patterns capture April-to-May ENSO decay and an opposing Indian Ocean adjustment, consistent with coupled Indo-Pacific variability. During 2008-2022, the GTV index prediction outperforms Ni\~{n}o3.4-based and sparse-regression benchmarks. Further improvement is achieved by including interactions of May Indo-China Peninsula soil moisture with Indian Ocean SST, importantly also correcting several underestimated wet summers. Circulation diagnostics associate the learned index with a summer geopotential-height dipole and Yangtze moisture convergence, while dry-Indochina/warm-Indian-Ocean conditions coincide with stronger westward extension of the western Pacific subtropical high.

\newpage
\section*{Introduction}\label{Sec1: intro}
Summer precipitation over the middle and lower reaches of the Yangtze River basin exerts significant control on floods, water resources, agriculture, and energy systems across eastern China. In 2020, a persistent Meiyu season brought record-breaking rainfall and widespread flooding \citep{liu2020record}, while 2022 featured an unprecedented warm-season flash drought with rapid intensification \citep{wang2023high}. At seasonal time scales, these events reflect the compound influence of tropical-extratropical teleconnections and the regional circulation systems that transport moisture into East Asia, including the Western Pacific Subtropical High (WPSH). Under climate change, the interannual variability of dominant circulation systems is projected to strengthen, raising concern for future shifts in flood and drought risk \citep{chen2016intensification,yang2022increased}.

Seasonal prediction of Yangtze summer rainfall has long remained difficult because it is controlled by interacting processes across ocean-land regions and monsoon circulations. State-of-the-art dynamical seasonal forecast systems can provide useful guidance for the basin, as demonstrated by real-time and retrospective predictions of extreme summers such as 2016 and 2020 and by evidence that components of the East Asian summer monsoon are predictable from earlier initializations \citep{jiang2013seasonal,bett2018seasonal,ying2022seasonal,bett2021seasonal}. Their skill, however, generally decreases with forecast lead time because of model errors and uncertainty in initial and boundary conditions. Observation-based prediction therefore often exploits sea surface temperature (SST), and compresses the climate state into indices such as Ni\~{n}o3.4 or a small number of leading modes \citep{huang2022skillful}. Such approaches can miss important information because the dominant oceanic control on East Asian summer rainfall has shifted in recent decades \citep{le2023climate,vos2026climate}. Specifically, eastern China summer rainfall now depends more critically on the rate of El Ni\~{n}o-Southern Oscillation (ENSO) decay \citep{jiang2019northwest} and the coupled Indo-western Pacific response \citep{liang2026statistical} than the ENSO amplitude alone \citep{sun2021changing,yu2025weakened}. The relevant precursor may therefore reside in the spatially heterogeneous and month-to-month evolution of spring SST across multiple ocean basins rather than in one or two static indices \citep{liang2023prediction}. 

Land-surface memory provides a second potential source of seasonal predictability \citep{gao2020interdecadal,gao2020effect,hsu2003relationship,xu2024potential}. In particular, spring soil moisture over the Indo-China Peninsula (ICP) can influence subsequent land heating and the western Pacific circulation, with its impact depending strongly on the surrounding oceanic state. During the record-breaking 2020 Meiyu season, for example, dry ICP soil moisture and a warm Indian Ocean acted synergistically to enhance ICP surface heating, favor westward extension of the WPSH, and increase Yangtze rainfall \citep{dong2024synergistic}. This suggests that land information may complement SST-based prediction, but its contribution may arise less from soil moisture alone than from state-dependent land-ocean interactions. Distinguishing these effects is therefore important for determining whether land-surface information provides additional seasonal predictability beyond the large-scale SST signal.

In this study, we quantitatively assess whether the spatiotemporal evolution of spring SST provides additional predictability for Yangtze summer precipitation and whether land-surface information further improves seasonal prediction. We apply graph total variation (GTV) with temporal weighting to March-May (MAM) SST fields over the western Pacific-Australasian, Ni\~{n}o3.4, and tropical Indian Ocean sectors. GTV jointly imposes sparsity and graph-based smoothness in the predictors while learning spatially and temporally distributed SST coefficients, whose projection onto the observed SST field defines a scalar prediction index. Predictive skill is evaluated using 1961-2007 for training and 2008-2022 as a fully independent testing period. We further examine whether May Indo-China Peninsula soil moisture contributes additional information beyond SST, either independently or through interactions with ENSO and Indian Ocean SST. This framework allows us to test whether seasonal predictability arises primarily from evolving oceanic precursor patterns or is further modulated by state-dependent land-ocean interactions.

\begin{figure}[!ht]
\vspace{-2mm}
\centering\includegraphics[width=1.0\linewidth]{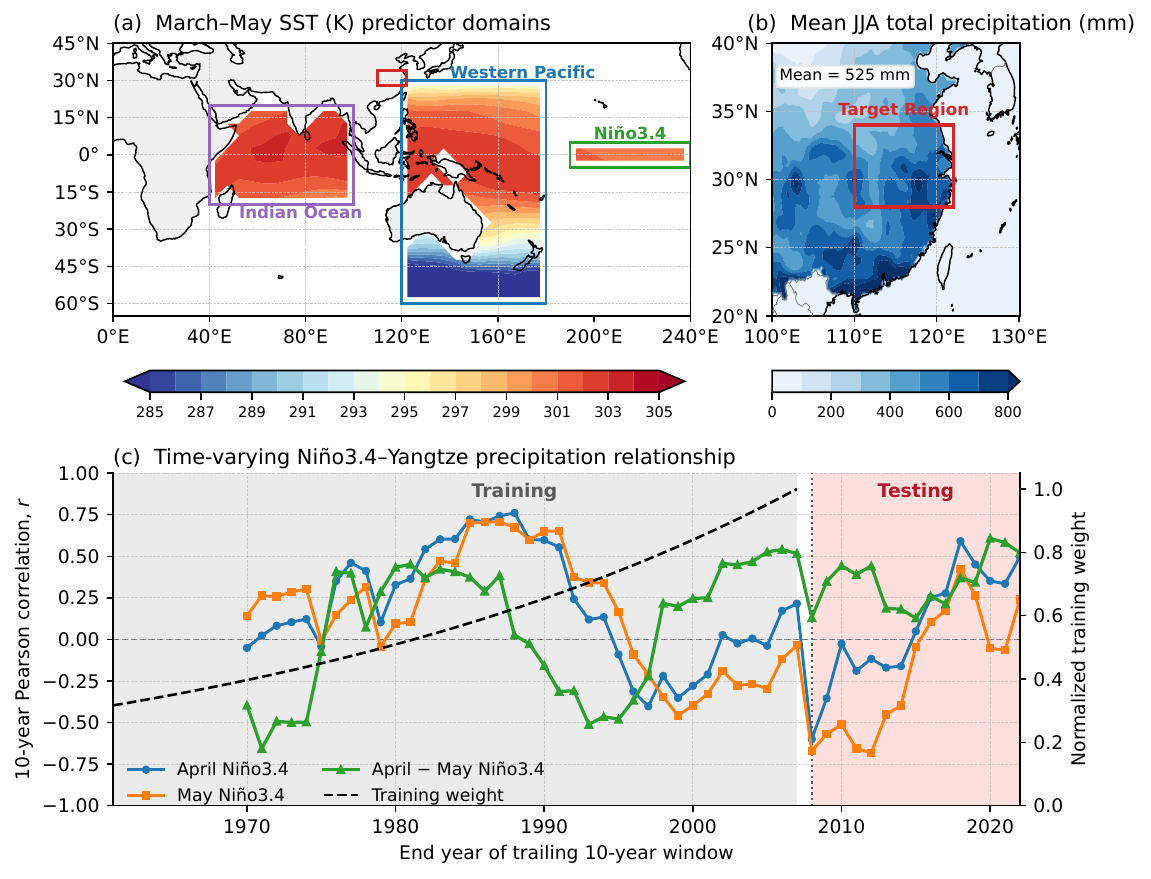}
\caption{
Predictor domains, target region, and nonstationarity diagnostics.
(a) Mean spring SST during the testing period (2008-2022), with rectangles indicating the three predictor domains: the extended western Pacific-Australasian sector, the Ni\~{n}o3.4 region, and the tropical Indian Ocean.
(b) Mean JJA precipitation during the testing period, with the red rectangle marking the target region over the middle and lower reaches of the Yangtze River.
(c) Ten-year moving correlations between Yangtze JJA precipitation and April Ni\~{n}o3.4 SST, May Ni\~{n}o3.4 SST, and their month-to-month change, $\Delta\mathrm{N3.4}=\mathrm{N3.4}_{\text{Apr}}-\mathrm{N3.4}_{\text{May}}$. Positive $\Delta\mathrm{N3.4}$ denotes cooling from April to May. The dashed black curve shows the normalized temporal weights used in GTV to emphasize more recent training years and account for nonstationary SST-precipitation relationships. \vspace{-3mm}}
\label{fig:1}
\end{figure}

\section*{Results}\label{Sec: Results}
\subsection*{Nonstationarity, GTV Performance, and Robustness}
\begin{figure}[!ht]
\vspace{-2mm}
\centering\includegraphics[width=1.0\linewidth]{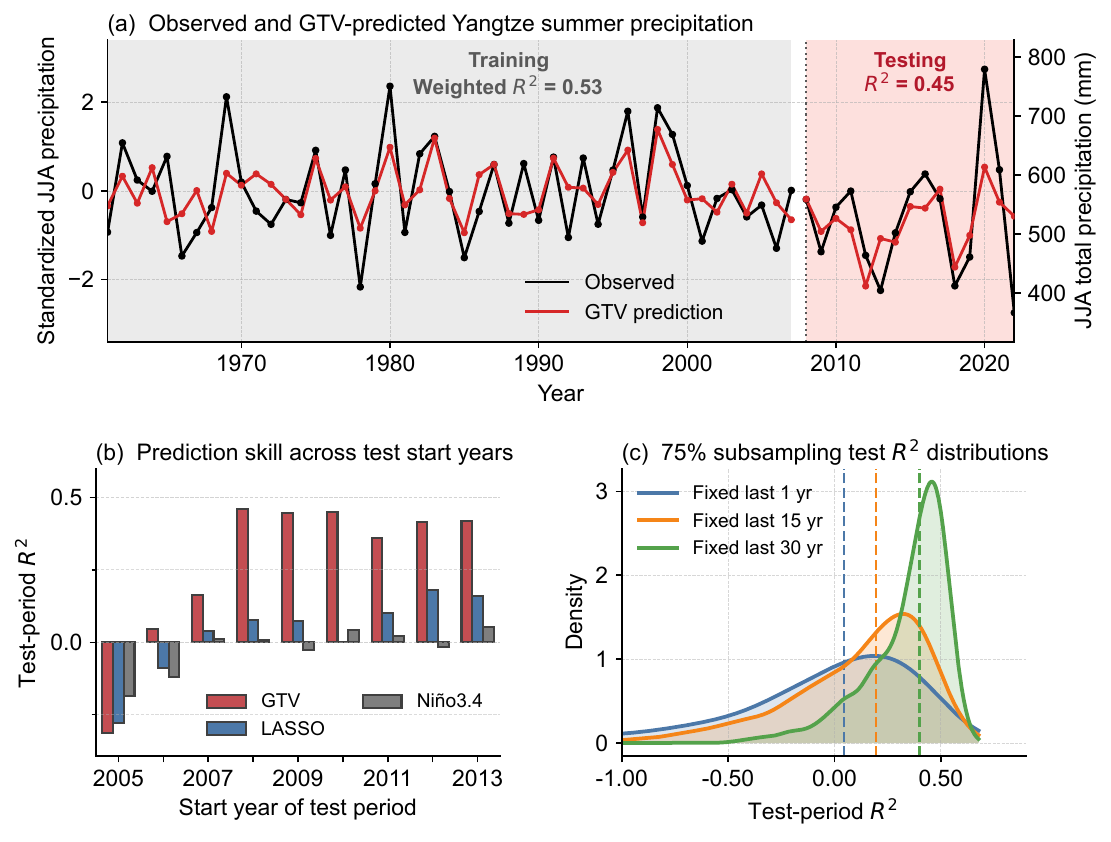}
\caption{
Prediction performance, benchmark comparison, and robustness of GTV.
(a) Observed and GTV-predicted Yangtze JJA precipitation during the 1961-2007 training period and the 2008-2022 testing period.
(b) Testing period $R^{2}$ for GTV (red), LASSO (blue), and Ni\~{n}o3.4 regression (gray) as a function of testing-period length. For each experiment, the most recent years are reserved for testing and all preceding years are used for training.
(c) Distributions of GTV testing $R^{2}$ from repeated training-period subsampling while forcing the most recent 1 (blue), 15 (orange), and 30 (green) training years to be retained. Vertical dashed lines denote the median. The 2008-2022 testing period is used across all realizations. \vspace{-3mm}}
\label{fig:2}
\end{figure}

Figure 1 summarizes the prediction setting and the nonstationarity motivating the temporally weighted GTV framework. The prediction uses MAM SST information from the western Pacific-Australasian, Ni\~{n}o3.4, and tropical Indian Ocean sectors to predict June-August (JJA) precipitation over the middle and lower Yangtze River. The April- and May-based Ni\~{n}o3.4 relationships vary substantially across decades, whereas the predictive relationship based on their difference, $\Delta\mathrm{N3.4}$, is comparatively more stable, particularly in recent decades (Figure \ref{fig:1}c).
Figure \ref{fig:2}a compares observed and GTV-predicted Yangtze JJA precipitation over the 1961-2007 training period and the independent 2008-2022 testing period. For display, the predictions are transformed to precipitation units using the training-period mean, standard deviation, and fitted linear trend. GTV closely follows the observed interannual variability, particularly during the later training period, yielding a weighted training $R^{2}$ of 0.53. More importantly, the model retains substantial skill during the fully independent testing period, with $R^{2}=0.45$. Most dry and wet summers are reproduced reasonably well, although several wet extremes, including 2016 and 2020, are underestimated in magnitude. This suggests that MAM SST evolution across the three predictor domains contains useful information about the subsequent Yangtze summer precipitation response, as further discussed in the next section. One notable exception is 2022, when GTV predicts a wet summer despite the observed flash drought. This mismatch is consistent with the importance of mid-to-high-latitude forcing and land-atmosphere coupling not represented by spring SST alone \citep{tang2023linkages,zhou2024impact,zuo2025predictability}. These remaining errors motivate the subsequent examination of land-surface information as an additional source of seasonal predictability.

Figure \ref{fig:2}b compares GTV with least absolute shrinkage and selection operator (LASSO) and a Ni\~{n}o3.4 regression while varying the length of the independent testing period. The Ni\~{n}o3.4 benchmark uses March-May monthly Ni\~{n}o3.4 SST as predictors, whereas LASSO is applied to the same spatially distributed SST information as GTV index. GTV provides the highest testing skill for nearly all train-test splits, except for the experiment with testing beginning in 2005. When the testing period is 15 years or shorter (i.e., begins in 2008 or later), GTV skill is relatively stable, with $R^{2}$ between approximately 0.38 and 0.45. In comparison, LASSO yields $R^{2}$ of about 0.05-0.20, while the Ni\~{n}o3.4 regression remains below 0.1. The improvement over LASSO indicates that the GTV advantage cannot be attributed to sparsity alone, but also reflects the graph constraint that preserves coherent spatiotemporal SST structures. When the testing period exceeds 15 years, however, the performance of all three methods decreases markedly. This behavior is consistent with the nonstationary motivation of the experimental design, suggesting that recent training samples better represent the contemporary SST-Yangtze rainfall relationship. GTV also performs well $(R^{2}=0.45, r=0.72)$ compared to previous Yangtze seasonal prediction studies, although direct numerical comparison is limited by differences in target region, forecast lead, and validation design. For example, dynamical hindcasts with GloSea5 achieved $r=0.55$ \citep{li2016skillful}, while a Random Forest model obtained a cross-validated $r=0.47$ \citep{he2021}. SST-based regression has achieved correlations up to $r=0.75$ \citep{qian2020seasonal}, while a statistical model repeatedly recalibrated over a moving 20-year training window achieved $r=0.76$ \citep{guo2017moving}. More recently, a physical-statistical model resolving the spatiotemporal evolution of continental East Asian rainfall achieved independent correlations of $0.70$-$0.75$ for its leading predictable modes \citep{ma2025skillful}. Although the predictands and validation designs differ across studies, these comparisons place GTV among the more skillful approaches.

The subsampling experiment in Figure \ref{fig:2}c provides a more direct test of the importance of recent training information while keeping the 2008-2022 testing period fixed. When only the most recent training year is kept during the subsampling, the median testing $R^{2}$ is approximately 0.05. The median increases to about 0.20 when the most recent 15 years are retained and to about 0.40 when the most recent 30 years are retained. These results indicate that consistently including recent observations improves prediction skill for the contemporary testing period, although the narrowing skill distributions partly reflect the reduced variability in training-sample composition as $K$ increases. Together with the train-test sensitivity experiment, this result indicates that recent observations contain particularly relevant information for the contemporary SST-Yangtze precipitation relationship and supports the use of temporal weighting in GTV. This behavior is consistent with documented multidecadal changes in ENSO-eastern China rainfall teleconnections \citep{sun2021changing,le2023climate,yu2025weakened} and with the dependence of East Asian summer circulation on the evolution of spring ENSO and the coupled Indo-western Pacific SST state \citep{jiang2019northwest,liang2026statistical}. 

\subsection*{GTV-Learned SST Evolution and Circulation Signature}

\begin{figure}[!ht]
\vspace{-2mm}
\centering
\includegraphics[width=1.0\linewidth]{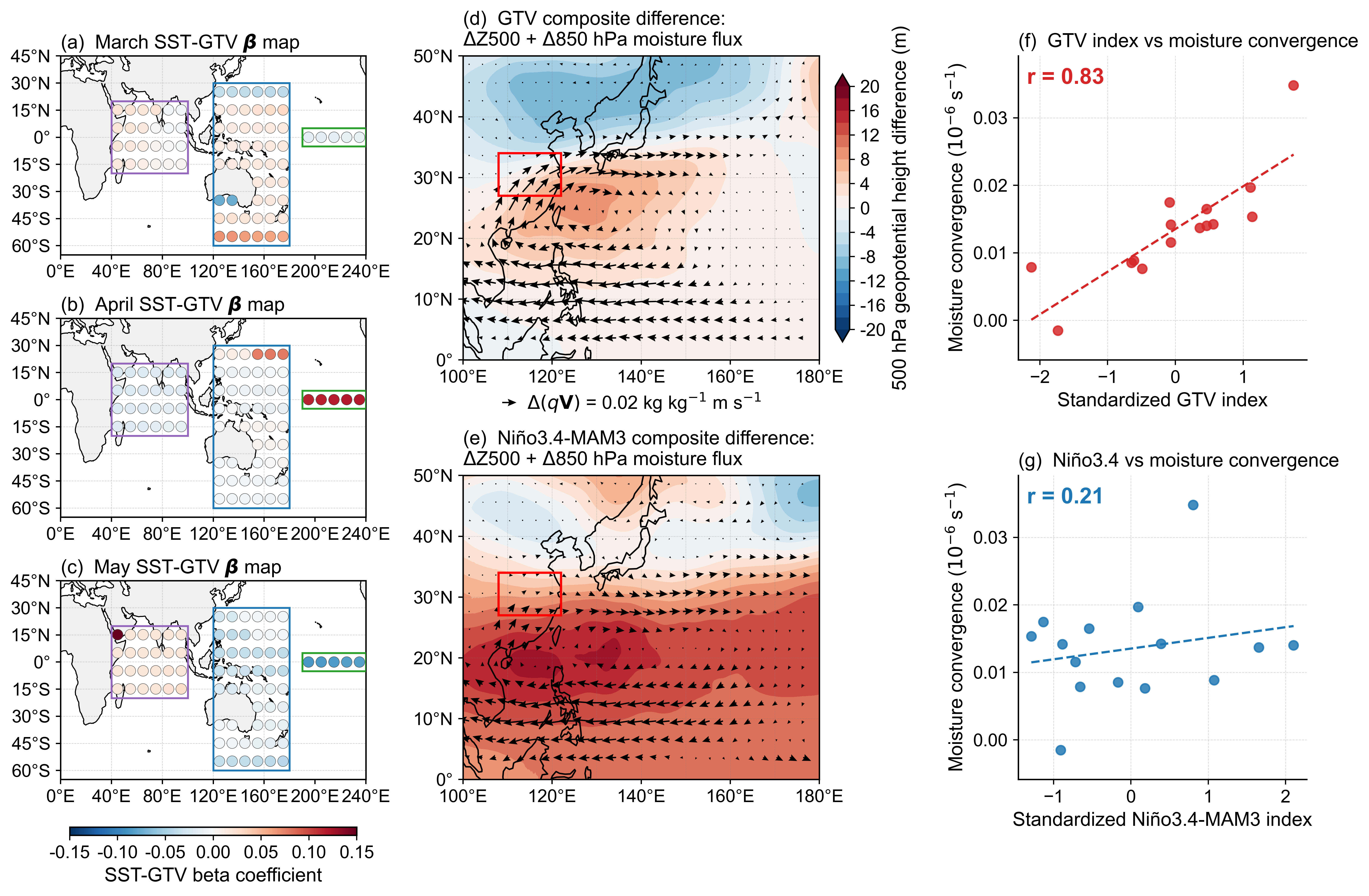}
\caption{
GTV-learned spring SST structure and its associated summer circulation and moisture-transport signatures.
(a-c) GTV regression coefficients ($\bm{\upbeta}$) for March, April, and May SST, respectively.
(d) High-minus-low composite differences in JJA 500-hPa geopotential height and 850-hPa moisture flux based on the GTV index over 1961-2022. Shading in (d,e) denotes 500-hPa geopotential-height differences, black vectors denote anomalous 850-hPa moisture-flux differences, $\Delta(q\mathbf{V})$, and the red rectangle marks the Yangtze target region. 
(e) Same as (d), but based on the Ni\~{n}o3.4 MAM regression index.
(f) Relationship between the GTV index and target-region moisture convergence over the testing period (2008-2022).
(g) Same as (f), but for the Ni\~{n}o3.4 MAM regression index.
\vspace{-3mm}}
\label{fig:3}
\end{figure}

The predictive advantage of GTV raises the question of whether its spatially distributed SST coefficients represent a physically meaningful precursor rather than statistical fit alone. The MAM coefficient maps reveal several coherent and evolving features (Figure \ref{fig:3}a-c). Most prominently, the Ni\~{n}o3.4 sector changes from positive coefficients in April to predominantly negative coefficients in May, indicating that GTV extracts information from the evolution of ENSO rather than from its amplitude in a single month. For example, a positive Ni\~{n}o3.4 SST anomaly in April followed by a negative anomaly in May contributes positively through both April-May coefficients. Built on this SST background, a pronounced spring cooling projects onto a stronger positive GTV index and, consequently, wetter Yangtze summer conditions. This evolving Pacific structure is consistent with evidence that the subsequent East Asian summer circulation depends strongly on the pace of ENSO decay and the associated western North Pacific response \citep{jiang2019northwest,jiang2023diverse,zhu2025physics}.

The Indian Ocean contribution becomes particularly evident in May, with the largest loadings centered over the Gulf of Aden and the adjacent western Indian Ocean. The pronounced May coefficients over the northwestern Indian Ocean suggest that GTV relates to part of the coupled Indo-western Pacific adjustment during late spring rather than a uniform basin-wide warming mode. This is consistent with the role of Indian Ocean SST anomalies in modulating convection, the western North Pacific circulation, and moisture transport toward East Asia \citep{wu2024role,dasgupta2026frequent}. A further feature is the positive March coefficients south and southeast of Australia, followed in April by positive coefficients over the northern Western Pacific. This sequence is suggestive of an evolving pattern consistent with a previously reported interhemispheric pathway linking SST variability near Australia and New Zealand to subsequent western North Pacific and East Asian circulation anomalies \citep{zhou2011sea,mamalakis2018new,cai2024impact}. The March signal may therefore represent an early extratropical component that evolves into the tropical and western North Pacific SST pattern emphasized later in spring. The stability of these learned patterns was further assessed through 200 temporally stratified training-data refits (Figure S1). The refitted coefficient maps show median spatial correlations of 0.63, 0.78, and 0.86 with the original March, April, and May maps, respectively (Figure S1j), while the positive-April/negative-May Ni\~{n}o3.4 coefficient contrast persists in 81\% of refits (Figure S1k). Despite variability in individual coefficient magnitudes (Figure S1d-f), these results support the relative stability of the broader GTV-learned SST patterns and their month-to-month evolution. 

The atmospheric circulation associated with this learned SST pattern further supports its physical interpretation. The composite difference between years in the upper and lower quartiles of the GTV index exhibits a pronounced meridional contrast in JJA 500-hPa geopotential height, with positive anomalies south of the Yangtze region and negative anomalies to the north (Figure \ref{fig:3}d). The Yangtze lies near the transition between these centers, accompanied by organized anomalous 850-hPa moisture transport toward the target region. In contrast, the Ni\~{n}o3.4-based composite emphasizes mainly the southern positive height anomaly, while the northern counterpart and associated moisture-flux response are weaker (Figure \ref{fig:3}e). Consistent with these circulation differences, the GTV index is strongly related to target-region moisture convergence during the testing period ($r=0.83$; Figure \ref{fig:3}f), whereas the corresponding relationship for the Ni\~{n}o3.4 index is much weaker ($r=0.21$; Figure \ref{fig:3}g). Thus, the distributed GTV coefficients appear to capture an evolving Indo-Pacific--South Pacific SST precursor that projects onto a more complete summer circulation and moisture-transport configuration relevant to Yangtze rainfall. This provides a physical basis for the additional predictive skill of GTV beyond a conventional Ni\~{n}o3.4 regression.

\subsection*{Land-Ocean Synergy as a State-Dependent Correction to GTV}\label{Sec: land-ocean}
\begin{figure}[!ht]
\vspace{-2mm}
\centering
\includegraphics[width=1.0\linewidth]{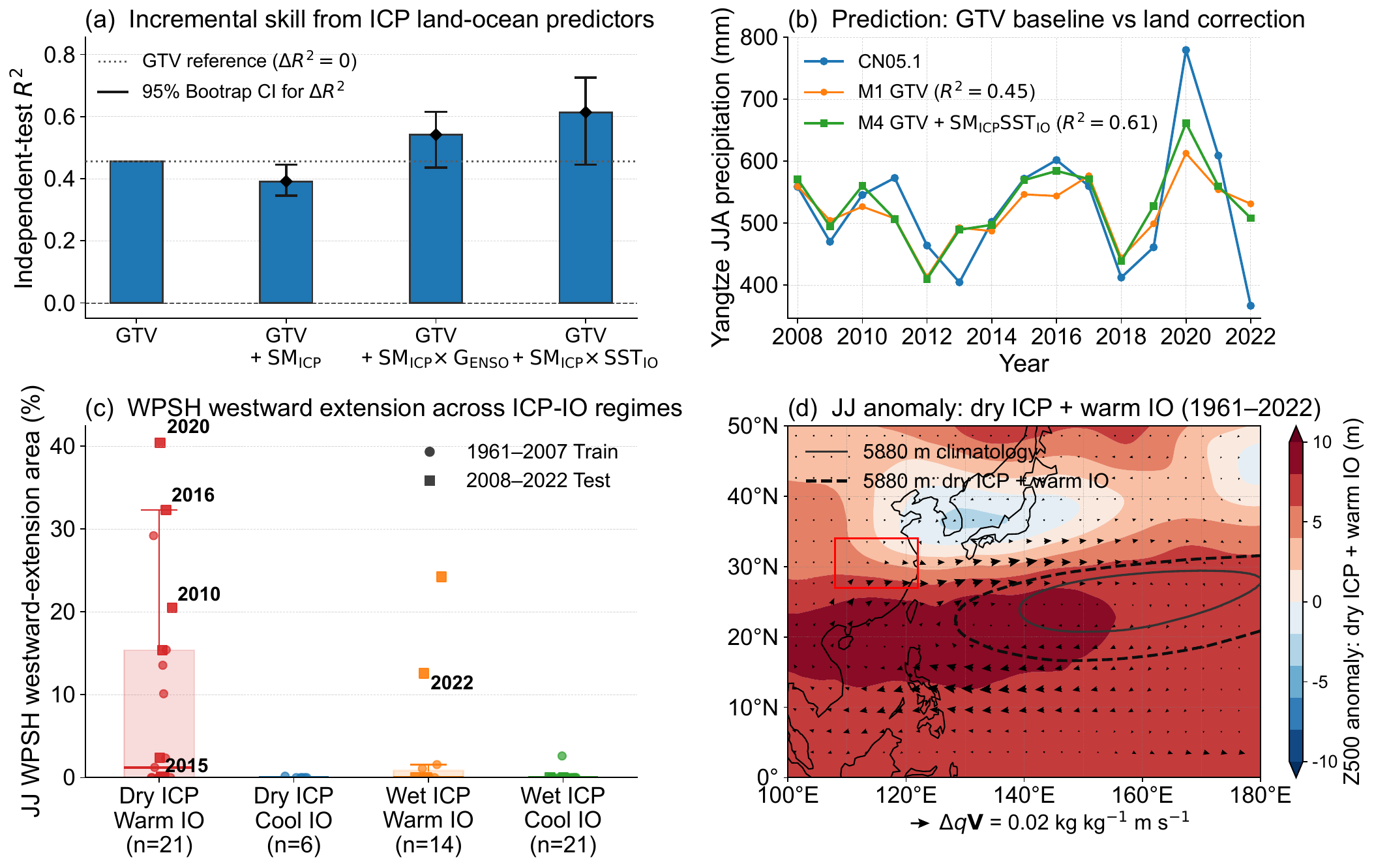}
\caption{
Land-ocean synergy and its contribution to Yangtze summer rainfall prediction.
(a) Test $R^{2}$ for the SST-only GTV model and models progressively incorporating May Indo-China Peninsula (ICP) soil moisture (SM), the SM$\times\text{G}_{\rm ENSO}$ interaction, and the SM$\times$Indian Ocean (IO) SST interaction. $\text{G}_{\rm ENSO}$ is a binary indicator of the preceding-winter ENSO background ($G_{\rm ENSO}=1$ when the December-February Ni\~{n}o3.4 SST anomaly is below a threshold $\tau$, and 0 otherwise). Error bars indicate 95\% bootstrap confidence intervals (CI) for the improvement relative to GTV ($\Delta R^2$; 20,000 resamples).
(b) Observed JJA precipitation and predictions from GTV and GTV+SM$\times$SST$_{\rm IO}$ during the 2008-2022 test period.
(c) June-July (JJ) western Pacific subtropical high (WPSH) westward-extension area (\%) for four ICP-SM/IO-SST regimes. Highlighted years indicate selected summers discussed in the text.
(d) Difference in JJ 500-hPa geopotential height (shading) and moisture flux (vectors) between dry-ICP/warm-IO years and the 1961-2022 climatology. Shading denotes 500-hPa geopotential-height anomalies, vectors denote 850-hPa moisture-flux anomalies, and contours show the 5880-m geopotential-height boundary. The red rectangle marks the Yangtze target region.
\vspace{-3mm}}
\label{fig:4}
\end{figure}

Although GTV captures the dominant SST-related predictability, its remaining errors suggest that land-surface conditions may provide additional information. Figure \ref{fig:4}a compares the GTV baseline with several models that additionally incorporate land-surface information. Adding May Indo-China Peninsula (ICP) soil moisture (SM) alone does not improve the prediction, with $R^2$ decreasing from 0.45 to about 0.40. In contrast, combining ICP SM with the large-scale oceanic state does improve the prediction: the SM$\times\text{G}_{\rm ENSO}$ interaction increases $R^2$ to 0.54, while the SM$\times$Indian Ocean (IO) SST interaction gives the largest improvement, reaching $R^2=0.61$. Bootstrap resampling of the 15 test summers (20,000 repetitions with replacement) yielded a 95\% confidence interval of $[-0.01, 0.27]$ for $\Delta R^2$. As this interval includes a negative lower bound, the improvement is not statistically significant at the 5\% level. However, we should interpret this lack of significance cautiously provided the limited test sample, which introduces considerable uncertainty into the bootstrap estimates. To assess the influence of 2020 (shown in panel (b)), we excluded that year without refitting the models. The improvement remained positive ($\Delta R^2=0.12$). Thus, the contrast between ICP SM alone and its ocean-dependent interactions highlights the importance of the background oceanic state. We focus below on the SM$\times$SST$_{\rm IO}$ interaction, examining its contribution to correcting underestimated wet years and its association with summer circulation anomalies.

Figure~\ref{fig:4}b shows that the SST-only GTV model underestimates several anomalously wet summers, particularly 2010, 2015, 2016, and 2020, whereas adding the SM$\times$SST$_{\rm IO}$ term reduces these errors. The correction is especially clear in 2020, when dry ICP conditions and a warm IO were shown to act together to enhance surface heating over the ICP, favor westward extension of the western Pacific subtropical high (WPSH), and increase Yangtze rainfall \citep{dong2024synergistic}. The four ICP-SM/IO-SST regimes further illustrate the state-dependent improvement (Figure~\ref{fig:4}c). For both the training and testing periods, the dry-ICP/warm-IO regime has the largest westward-extension area of the WPSH and includes several of the wet summers receiving strong positive corrections. This land-ocean synergy has been documented particularly during the June-July Meiyu season \citep{dong2024synergistic,gao2020effect}, when dry May ICP soil moisture enhances land-surface heating under a warm IO background and favors westward extension of the WPSH. Physically, dry ICP soil reduces evapotranspiration and enhances sensible heating, while warm IO SST favors an anomalous anticyclonic response over the western North Pacific. Their combined influence can therefore reinforce the westward extension of the WPSH \citep{xie2009,gao2020effect,gao2020interdecadal}.

The 2022 summer exhibits a useful contrast. Unlike the wet correction years, 2022 occurred under a wet-ICP/warm-IO state, and the SM$\times$SST$_{\rm IO}$ term reduces the wet bias of the SST-only GTV prediction. With the IO remaining warm but the ICP SM anomaly changing sign, the land-ocean interaction contributes in the opposite direction. Wetter ICP conditions favor greater evapotranspiration and weaker sensible heating, which may reduce the land-surface contribution to WPSH westward extension. Thus, the opposite corrections in 2020 and 2022 are consistent with the state-dependent nature of the SM-IO interaction rather than a simple additive effect of soil moisture.

The circulation anomalies associated with the dry-ICP/warm-IO regime provide further support for this interpretation (Figure~\ref{fig:4}d).  The dry-ICP/warm-IO regime is associated with positive JJ 500-hPa geopotential-height anomalies over the western North Pacific and a clear westward displacement of the 5880-m contour, consistent with an enhanced westward extension of the WPSH. The moisture-flux anomalies are weaker than those associated with the GTV index in Figure~\ref{fig:3}, suggesting that the SM-IO interaction primarily modifies the position and extent of the subtropical circulation rather than generating a separate large-scale moisture-transport signal. ICP SM therefore acts as a state-dependent correction to the dominant SST-driven circulation captured by GTV, helping account for years in which the SST-only model under- or overestimates the Yangtze rainfall response.

\section*{Discussion}\label{Sec: Dis}
Spring SST evolution provides substantial and physically interpretable predictability for Yangtze summer precipitation. The GTV index achieves a predictive skill of $R^{2}=0.45$ ($r=0.72$) during 2008-2022 and outperforms Ni\~{n}o3.4 regression and LASSO across most train-test splits. Prediction skill also becomes more stable when recent observations are retained, consistent with documented multidecadal changes in ENSO-East Asian rainfall relationships \citep{sun2021changing,yu2025weakened}. Importantly, the learned SST pattern is physically coherent rather than purely statistical, capturing April-to-May ENSO evolution, a strengthening Indian Ocean signal in late spring, and an earlier South Pacific-western Pacific signal. The resulting GTV index is associated with a clearer meridional 500-hPa geopotential-height contrast and stronger moisture convergence over the Yangtze region than the Ni\~{n}o3.4-based index, providing a physical basis for its additional prediction skill.

May Indo-China Peninsula (ICP) soil moisture alone provides little added skill beyond GTV, whereas its interaction with ENSO and, especially, Indian Ocean (IO) SST increases the $R^{2}$ from 0.45 to 0.54 and 0.61, respectively. The synergistic SM$\times$SST$_{\rm IO}$ term corrects several wet summers underestimated by GTV, with the dry-ICP/warm-IO state emerging as the dominant regime for positive corrections. This regime is also associated with a stronger westward extension of the western Pacific subtropical high, consistent with previously reported land-ocean synergy during the Meiyu season \citep{gao2020effect,gao2020interdecadal,dong2024synergistic}. These results indicate that ICP soil moisture contributes primarily through a state-dependent interaction with the oceanic background rather than as an independent predictor.

Some limitations remain. The nonstationarity means that the learned SST-rainfall relationship may change in the future. Temporal weighting assumes that recent observations provide more relevant information about the near-future relationship, but cannot anticipate changes that differ from those observed historically. This motivates periodically updating the learned relationship as new observations become available. Exceptional events involving strong internal atmospheric variability, mid-to-high-latitude forcing, or land--atmosphere feedbacks may also not be fully captured by spring SST alone, as illustrated by the 2022 flash drought \citep{tang2023linkages,zhou2024impact,zuo2025predictability}.  Future work could project SST forecasts from multi-model seasonal prediction ensembles onto the learned GTV pattern, yielding an ensemble of GTV indices that could be used to calibrate or merge rainfall forecasts and quantify prediction uncertainty. The proposed approach could also be extended toward spatially resolved and subseasonal prediction by incorporating additional land-surface predictors, such as regional soil moisture and land-surface temperature, together with intraseasonal information \citep{strazzo2019application,miller2021skillful}.  Overall, the results show that explicitly representing spring SST evolution and its interaction with land conditions provides a skillful and physically interpretable framework for Yangtze summer rainfall prediction.

\section*{Methods}\label{Sec: data}
\subsection*{Data and prediction target}
The precipitation data are obtained from the gauge-based CN05.1 gridded observational dataset (1961-2022) at $0.25^\circ \times 0.25^\circ$ resolution \citep{wu2013}. The predictand is area-mean JJA accumulated precipitation over $110^\circ$-$122^\circ$E and $28^\circ$-$34^\circ$N, representing the middle and lower Yangtze River and primary Meiyu region. The monthly CN05.1 precipitation rates (mm day$^{-1}$) are converted to monthly totals by multiplying by the number of days in each month, summed over June-August at each grid cell, and then spatially averaged over the target region using cosine-latitude weighting, yielding JJA precipitation in mm season$^{-1}$. Monthly SST fields are taken from ERA5 at the same resolution. Predictors are MAM SSTs from three Indo-Pacific sectors: the extended western Pacific-Australasian sector ($120^\circ$-$180^\circ$E, $60^\circ$S-$30^\circ$N), Ni\~{n}o3.4 ($190^\circ$-$240^\circ$E, $5^\circ$S-$5^\circ$N), and the tropical Indian Ocean ($30^\circ$-$100^\circ$E, $20^\circ$S-$20^\circ$N). These domains capture major tropical SST sources of seasonal predictability for subsequent Yangtze summer rainfall, together with potential extratropical South Pacific precursors. Previous studies indicate that spring SST anomalies across these regions can persist or evolve into summer and modulate circulation conditions affecting Yangtze rainfall \citep{zhou2011sea,ha2019relationship,zhang2022indian,zhou2021historic,wen2022response}. 

Figure \ref{fig:1}a-b shows MAM SST over the predictor domain and mean JJA precipitation over the target region during the testing period. Before fitting, SST is aggregated to a $10^\circ \times 10^\circ$ grid to emphasize large-scale evolution while limiting predictor dimensionality. The 1961-2022 record is divided into a training period of 1961-2007 and a fully independent testing period of 2008-2022. The detailed validation and model-selection procedures are described below. ERA5 850-hPa winds and specific humidity and 500-hPa geopotential height are additionally used to diagnose circulation and moisture-transport anomalies associated with the learned GTV index.

\subsection*{Prediction model and GTV regularization}
We formulate the seasonal prediction problem as a linear regression between spring SST predictors and Yangtze summer precipitation. Let $y_t$ denote the standardized JJA precipitation over the target region in the middle and lower reaches of the Yangtze River in year $t$, and let $\mathbf{x}_t \in \mathbb{R}^{p}$ denote the vector of MAM SST predictors. The basic prediction model is
\begin{equation}
y_t = \mathbf{x}_t^\top \bm{\upbeta} + \varepsilon_t ,
\label{eq:linear_model}
\end{equation}
where $\bm{\upbeta}$ is the coefficient vector and $\varepsilon_t$ is the residual. The learned GTV-based SST evolution index is then defined as
\begin{equation}
I_t^{\mathrm{GTV}} = \mathbf{x}_t^\top \widehat{\bm{\upbeta}}.
\label{eq:gtv_index}
\end{equation}
Here, $\widehat{\bm{\upbeta}}$ represents the spatially and temporally distributed SST weights learned by GTV, whereas $I_t^{\mathrm{GTV}}$ is the resulting scalar index obtained by projecting the SST field in year $t$ onto those weights. 

Because SST-eastern China rainfall teleconnections exhibit pronounced decadal variability \citep{sun2021changing,yu2025weakened,vos2026climate}, we use temporal weighting to reduce the influence of historical relationships that may be less representative of the contemporary climate. The moving-correlation diagnostic used to illustrate this nonstationarity is shown in Figure \ref{fig:1}c.

We therefore assign larger weights to more recent training years, so that the fitted model is guided more strongly by the contemporary teleconnection pattern while still retaining information from the full training record. For a training period of length $n_{\mathrm{train}}$, the weight assigned to year $t$ is
\begin{equation}
w_t =
\frac{
n_{\mathrm{train}}\exp\left[\gamma\left(t-n_{\mathrm{train}}\right)\right]
}{
\sum_{s=1}^{n_{\mathrm{train}}}
\exp\left[\gamma\left(s-n_{\mathrm{train}}\right)\right]
},
\qquad t=1,\ldots,n_{\mathrm{train}},
\label{eq:sample_weight}
\end{equation}
where $\gamma$ controls the strength of the temporal weighting. The weighted predictor matrix and target vector are given by
\begin{equation}
\widetilde{\mathbf{X}} = \mathbf{W}^{1/2}\mathbf{X},
\qquad
\widetilde{\mathbf{y}} = \mathbf{W}^{1/2}\mathbf{y},
\label{eq:weighted_data}
\end{equation}
where $\mathbf{W}=\mathrm{diag}(w_1,\ldots,w_{n_{\mathrm{train}}})$. We set $\gamma=0.025$. A sensitivity analysis over $\gamma=0.01$-$0.05$ yielded similar prediction skill, with testing $R^{2}$ ranging from 0.41 to 0.45.

After aggregation to the $10^\circ \times 10^\circ$ grid, the MAM SST predictor field contains $p=261$ predictors, exceeding the 47 training years (1961-2007). This $p>n_{\mathrm{train}}$ setting makes ordinary regression underdetermined. We therefore use GTV regularization to jointly impose sparsity and graph-based smoothness, allowing the model to select a compact set of informative SST predictors while preserving coherent SST-evolution structures across dynamically related grid cells. We estimate the coefficient vector by solving the following GTV-regularized optimization problem:
\begin{equation}
\widehat{\bm{\upbeta}}
=
\underset{\bm{\upbeta}}{\arg\min}
\left\{
\frac{1}{2}
\left\|
\widetilde{\mathbf{y}}
-
\widetilde{\mathbf{X}}\bm{\upbeta}
\right\|_2^2
+
\lambda_1
\left\|
\bm{\upbeta}
\right\|_1
+
\lambda_2
\sum_{(j,k)\in E}
|\Sigma_{jk}|^{1/2}
\left|
\beta_j
-
s_{jk}\beta_k
\right|
\right\},
\label{eq:gtv_objective}
\end{equation}
Here, $\lambda_1$ controls sparsity and $\lambda_2$ controls graph smoothness. The sparsity term $\|\bm{\upbeta}\|_1$ selects a compact set of informative SST predictors. The GTV term penalizes coefficient differences between connected SST predictors, with stronger penalties assigned to pairs that are more strongly correlated. Specifically, for each edge $(j,k)\in E$, the penalty is weighted by the square root of the covariance $|\Sigma_{jk}|^{1/2}$ and uses $s_{jk}=\mathrm{sign}(\Sigma_{jk})$ to account for whether the two predictors covary in phase or out of phase. Thus, positively correlated SST predictors are encouraged to have similar coefficients, whereas negatively correlated predictors are encouraged to have coefficients of opposite sign. 

\subsection*{Land-surface enhancement of SST prediction}
Land-surface memory provides additional seasonal predictability beyond SST forcing. Here we focus on May ICP soil moisture, which has a documented influence on Meiyu circulation and precipitation extremes \citep{gao2020interdecadal,gao2020effect,dong2024synergistic}. Unlike the high-dimensional SST fields used by GTV, the land correction is formulated using low-dimensional regional indices. Specifically, the ICP is defined as $95^\circ$-$110^\circ$E, $8^\circ$-$23^\circ$N, and the predictor is calculated as the cosine-latitude-weighted area-mean May surface soil moisture from ERA5. The large-scale oceanic background is summarized by scalar indices: a preceding-winter (DJF) Ni\~{n}o3.4 index and a March-May tropical Indian Ocean SST index, each calculated from regional-mean SST. We use these scalar predictors to model the residual of the SST-based GTV prediction. Three extensions are considered: GTV+$\text{SM}_{\rm ICP}$, GTV+$\text{SM}_{\rm ICP}\times \text{G}_{\rm ENSO}$, and GTV+$\text{SM}_{\rm ICP}\times\text{SST}_{\rm IO}$, where $\text{G}_{\rm ENSO}$ is a binary indicator of the preceding-winter ENSO background ($G_{\rm ENSO}=1$ when the magnitude of the DJF Ni\~{n}o3.4 SST anomaly is below a cross-validated threshold $\tau$ (selected between 0.5 and 1.0 K), and 0 otherwise), and $\mathrm{SST}_{\rm IO}$ is the MAM tropical Indian Ocean SST index. The latter two extensions test whether the contribution of ICP soil moisture depends on the large-scale oceanic state rather than acting independently. 

Since temporal weighting can make the in-sample residuals from the GTV model fitted to the full training period artificially small in later training years, we train the land corrections using nested blocked out-of-fold (OOF) GTV residuals \citep{roberts2017cross}. The 1961-2007 training period is divided into five contiguous outer blocks. For each held-out block, $\lambda_1$ and $\lambda_2$ are reselected from the remaining years using four-fold blocked cross-validation before generating the OOF GTV predictions. The temporal-decay parameter is fixed at $\gamma=0.025$ throughout. The held-out predictions from the five outer folds are then combined to form
\begin{equation}
r_t^{\rm OOF}
=
y_t-\widehat{y}_{t,\rm GTV}^{\rm OOF}.
\end{equation}
Thus, each residual is obtained from a GTV model that excludes the entire temporal block containing year $t$, rather than only that individual year. This provides a more realistic estimate of the SST-only prediction error and reduces overfitting of the subsequent land correction. For each extension, the land coefficient is estimated by weighted least squares,
\begin{equation}
\widehat{\bm{\beta}}_{\rm land}
=
\arg\min_{\bm{\beta}}
\sum_t w_t
\left(
r_t^{\rm OOF}
-
\mathbf{z}_t^\top\bm{\beta}
\right)^2,
\end{equation}
where $\mathbf{z}_t$ contains the corresponding land or land-SST interaction predictor.

For the physical diagnosis of the SM$\times$SST$_{\rm IO}$ interaction, years are classified into four ICP-SM/IO-SST regimes according to the signs of the standardized May ICP soil moisture and MAM Indian Ocean SST anomalies. Standardization is based on the 1961-2007 training-period mean and standard deviation, such that negative (positive) anomalies denote dry (wet) ICP conditions and cool (warm) IO conditions. 

\subsection*{Experimental design and validation}\label{Sec3: Experimental Design}
The experiments are designed to evaluate (i) the out-of-sample prediction skill and robustness of GTV relative to conventional SST-based and regularization benchmarks, (ii) the physical basis of the spatially distributed GTV SST coefficients through their associated circulation and moisture-transport patterns, and (iii) the added predictive value of ICP soil moisture and its underlying land-ocean interaction mechanisms. In the main experiment, the record is divided into 1961-2007 for model training and 2008-2022 for independent testing. During model fitting, precipitation and SST are detrended and standardized using training-period statistics, and the temporally varying sample weights defined by Equation \eqref{eq:sample_weight} assign progressively greater influence to recent training years.

The graph used in the GTV penalty follows \citet{stevens2021graph}. For the third term in Equation~\eqref{eq:gtv_objective}, edges are retained only when the absolute training-period correlation exceeds $\theta=0.5$, removing weakly related predictor pairs and emphasizing coherent SST-evolution patterns. The sparsity and graph-smoothness parameters, $\lambda_1$ and $\lambda_2$, are selected by five-fold blocked cross-validation, in which the training years are divided into five contiguous chronological blocks. Each block is held out in turn, and the remaining blocks are used for fitting. Precipitation detrending, precipitation and SST standardization, and SST graph construction are performed using the complete training period (1961-2007), with the resulting transformations and graph held fixed across the five cross-validation folds. Coefficients $\bm{\upbeta}$ are estimated using an alternating direction method of multipliers (ADMM) solver \citep{boyd2011distributed}. The final model is refit on the full weighted training set using the selected hyperparameters and then evaluated on the independent testing period. 

Two sensitivity experiments assess prediction robustness and the role of nonstationarity. First, the independent testing period is varied from 12 to 20 years to examine sensitivity to the train-test split. Second, repeated subsampling of the 47-year training record is performed while always retaining the most recent $K$ years ($K=1,15,30$), testing whether greater inclusion of recent observations stabilizes and improves prediction skill. GTV is further benchmarked against a Ni\~{n}o3.4 regression and LASSO applied to the same SST information, assessing whether the learned spatiotemporal SST pattern adds predictive value beyond a conventional ENSO index and sparsity alone.

\subsection*{Circulation and WPSH diagnostics}
Finally, we examine the physical interpretation of the learned GTV index. We compare high-minus-low composites of 500-hPa geopotential height and 850-hPa moisture flux based on the GTV index and the Ni\~{n}o3.4 index. This composite analysis is used to diagnose whether the GTV index preferentially captures circulation and moisture-inflow patterns that are dynamically favorable for Yangtze summer precipitation. To investigate land-associated physics, we follow the WPSH westward-extension-area (WEA) concept of \citet{li2024} and quantify the June-July WPSH WEA as the cosine-latitude-weighted percentage of the $105^\circ$-$130^\circ$E, $10^\circ$-$30^\circ$N sector with 500-hPa geopotential height exceeding 5880 m. Larger WEA therefore indicates a greater westward occupation of the WPSH over its western sector.

\section*{Data Availability}
The gauge-based CN05.1 gridded observational dataset (1961-2022) is described in \citet{wu2013} and available from Nansen-Zhu International Research Centre at \url{https://nzc.iap.ac.cn/content?cid=24&aid=999}. The ERA5 monthly reanalysis data are described in \citep{hersbach2020era5}. SST and soil moisture are obtained from the ERA5 monthly averaged data on single levels from 1940 to present, available at \url{https://cds.climate.copernicus.eu/datasets/reanalysis-era5-single-levels-monthly-means}. The 850-hPa winds and specific humidity and 500-hPa geopotential height used for the circulation diagnostics are obtained from the ERA5 monthly averaged data on pressure levels, available at \url{https://cds.climate.copernicus.eu/datasets/reanalysis-era5-pressure-levels-monthly-means}. 

\section*{Code Availability}
The GTV MATLAB processing scripts will be archived in Zenodo together with the postprocessed ERA5 SST and precipitation data, the learned GTV index time series, coefficient maps, benchmark results, and figure source data. An Excel sheet containing the preprocessed ERA5 SST predictors, Yangtze precipitation target, and GTV output time series is included in the supporting information for review.

\section*{Acknowledgments}
This study is supported by the National Natural Science Foundation of China (Grant No. 42688201).

\section*{Author contributions}
Y.G. performed the formal analysis, designed the experimental framework, prepared the figures, and wrote the original draft. H.C. contributed to the conceptualization of the study, provided resources, and proposed incorporating land-ocean synergy to improve predictive skill. R.L. contributed to data curation and preprocessing and helped design the subsampling experiments. E.F.-G. contributed to the conceptualization and experimental design and edited the manuscript. J.A.V. provided resources and edited the manuscript.

\section*{Competing interests}
The authors declare no competing interests

\printbibliography

\end{document}


\maketitle
\vspace{-1.5em}

\begin{flushleft}
\small
\textsuperscript{1} State Key Laboratory of Climate System Prediction and Risk Management/Key Laboratory of Meteorological Disaster, Ministry of Education/Collaborative Innovation Center on Forecast and Evaluation of Meteorological Disasters, Nanjing University of Information Science and Technology, Nanjing, Jiangsu, China.\par
\textsuperscript{2} School of Atmospheric Sciences, Nanjing University of Information Science and Technology, Nanjing, Jiangsu, China.\par
\textsuperscript{3} Department of Civil and Environmental Engineering, University of California, Irvine, California, USA.\par
\textsuperscript{4} Department of Earth System Science, University of California, Irvine, Irvine, California, USA.\par
\textsuperscript{5} Corresponding author: \href{mailto:haishan@nuist.edu.cn}{\textcolor{blue}{haishan@nuist.edu.cn}}.
\end{flushleft}

\vspace{0.8em}
\section*{Supplementary Figure(s)}

\begin{figure}[!htbp]
    \centering
    \includegraphics[width=\linewidth]{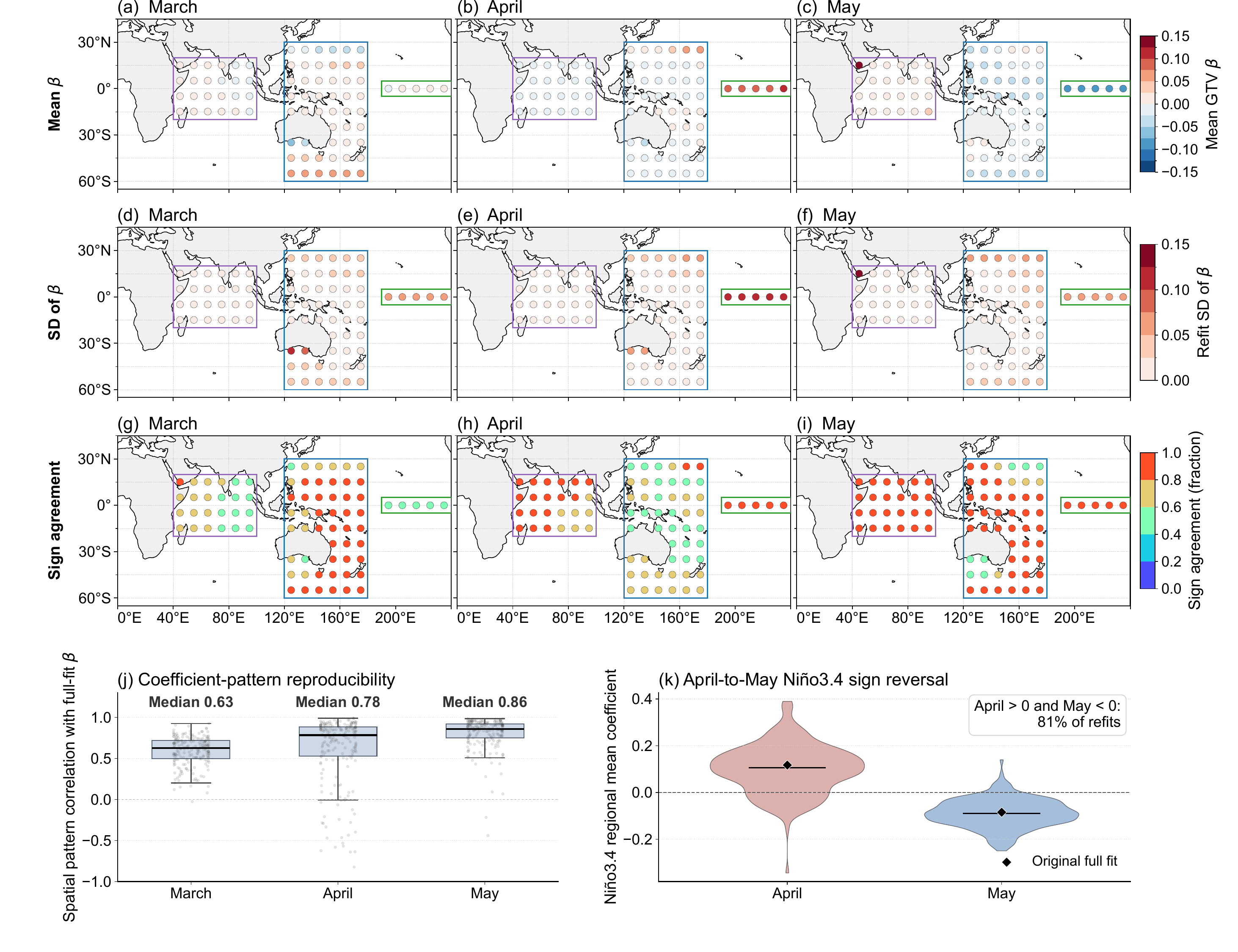}
    \caption{\textbf{Stability of GTV-learned spring SST coefficients under temporally stratified subsampling.}
    The GTV model was refitted 200 times using 38 of the 47 training years (1961-2007), sampled without replacement across five temporal strata: 1961-1970, 1971-1980, 1981-1990, 1991-2000, and 2001-2007. Eight years were sampled from each of the first four 10-year periods and six years from the final 7-year period, retaining approximately 80\% of the training years within each stratum. This design maintains representation across the training record and avoids subsamples disproportionately concentrated in earlier years, given potential nonstationarity in the SST-precipitation relationship. The original model-selected regularization parameters were held fixed, while the GTV graph was rebuilt for each subsample. (a-c) Mean regression coefficients ($\bm{\beta}$) across refits for March, April, and May, respectively. (d-f) Corresponding standard deviations (SD) of the coefficients. (g-i) Sign agreement, defined as the fraction of refits in which the coefficient at each grid cell retains the sign of the full-training fit. (j) Distribution of spatial pattern correlations between the refitted coefficient maps and the original full-fit maps shown in Figure~3a-c of the main text. (k) Distributions of Ni\~no3.4 regional-mean coefficients for April and May; black diamonds indicate the original full-fit coefficients. The positive-April/negative-May coefficient contrast occurs in 81\% of refits, supporting the reproducibility of this learned ENSO-evolution feature. The refit SDs quantify sensitivity to training-sample composition rather than formal confidence intervals.}
    \label{fig:S1}
\end{figure}